\documentclass[twocolumn,showpacs,preprintnumbers,amsmath,amssymb]{revtex4}

\usepackage{graphicx}
\usepackage{dcolumn}
\usepackage{bm}

\newcommand{\gsim}{\mbox{\raisebox{-0.6ex}{$\stackrel{>}{\sim}$}}\:}

\begin{document}

\preprint{YITP-06-05}

\title{Pentaquark state in pole-dominated QCD sum rules}

\author{Toru Kojo$^{1,2}$, Arata Hayashigaki$^3$, Daisuke Jido$^4$}

\affiliation{
$^1$Department of Physics, Kyoto University, Kyoto 606-8502, Japan \\
$^2$Department of Physics, Osaka University, Toyonaka 560-0043, Japan \\
$^3$Institut f\"ur Theoretische Physik, J.W. Goethe Universit\"at,
          D-60438 Frankfurt am Main, Germany \\
$^4$Yukawa Institute for Theoretical Physics, Kyoto University,
          Kyoto 606-8502, Japan
}

\date{\today}

\begin{abstract}
We propose a new approach in QCD sum rules applied for exotic hadrons with 
a number of quarks, exemplifying the pentaquark $\Theta^{+} (I=0,J=1/2)$
in the Borel sum rule. 
Our approach enables reliable extraction of the pentaquark properties
from the sum rule with good stability in a remarkably wide Borel window. 
The appearance of its valid window originates from a favorable setup of 
the correlation functions with the aid of {\it chirality} of the 
interpolating fields
on the analogy of the Weinberg sum rule for the vector currents.
Our setup leads to large suppression of the continuum contributions
which have spoiled the Borel stability in the previous analyses, 
and consequently enhances importance of 
the higher-dimensional contributions of the OPE, which 
are 
indispensable for investigating the pentaquark properties. 
Implementing the OPE analysis up to dimension 15,
we find that the sum rules for the chiral-even and odd parts independently
give the $\Theta^{+}$ mass of $1.68\pm 0.22$ GeV with uncertainties 
of the condensate values.
Our sum rule indeed gives rather flat Borel curves almost independent 
of the continuum thresholds both for the mass and pole residue.
Finally, we also discuss possible isolation of the observed states 
from the $KN$ scattering state on view of 
chiral symmetry. 

\end{abstract}


\maketitle

\section{Introduction}
The first discovery of the baryonic resonance with $S=1$,
$\Theta^{+}(1540)$, and its confirmation in subsequent low-energy 
exclusive experiments in 2003 \cite{exp}
triggered tremendous amount of theoretical works on 
exotic hadrons in a short time.
Many of their studies have been devoted to clarifying mainly 
its possible structure and its property such as spin and parity, 
and further searching for other exotic states \cite{theo,JW}.
So far, it is experimentally known that the $\Theta^+$ has 
minimal quark contents, $uudd\bar{s}$, from observation of 
its decay mode into $KN$, with $I_3=0$ and most likely 
an isospin singlet $I=0$ \cite{exprev}.

Yet the experimental evidence for the existence of $\Theta^{+}$ is 
not so obvious.
While new data with better statistics from the LEPS collaboration
consolidate their positive evidence of the $\Theta^+$ \cite{leps},
the most recent experiment in low-energy exclusive
reaction with high statistics by the CLAS Collaboration \cite{clas}, 
however, showed {\it negative} evidence for the $\Theta^+$, 
suggesting that their previous result would be 
just a statistical fluctuation.
The inclusive high-energy processes  
in $e^+e^-$ or hadron collisions 
have also claimed {\it no} evidence \cite{high_energy}. 
The disagreement between the LEPS and the other experiments 
would possibly originate from their differences of experimental setup 
and kinematical conditions; the former experiment covers well
the forward angle, where the $\Theta^+$ would be produced 
by meson-exchange production mechanism at low energy.

Theoretical study on existence of the $\Theta^{+}$ is also a very important 
issue. Investigation of such an exotic hadron can be the first step 
to explore the quark matter.
To identify the exotic state definitely, 
theoretical computations in direct approaches of QCD with 
less assumptions and better accuracy are getting more 
important. 
One of its possible analyses is the QCD sum rule (QSR) \cite{shifman}, 
which is a powerful tool to address 
directly nonperturbative dynamics peculiar to QCD 
as well as lattice QCD
and is a quite established approach 
for reproducing the baryon masses \cite{IoffeN}
including their resonance states \cite{Jido}.

Indeed, a number of the QSR analyses for scrutinizing 
the $\Theta^+$ mass were implemented with the help 
of the Borel sum rules (BSR's) \cite{SDO,bsr,Oga} 
and the finite energy sum rules \cite{fesr}.
It is generally known that the former technique 
is superior to the latter quantitatively,
because in the former the highly excited states can be controlled 
by the inverse Borel mass ($1/M$) to isolate the desired pole contribution. 
To the best of our knowledge, so far no BSR analyses for the $\Theta^{+}$ mass
have focused upon desirable {\it pole-dominance} of the $\Theta^+$,
even accounting for higher corrections 
of the operator product expansion (OPE).
But rather they have stuck with an undesirable continuum dominant region,
so that they could not establish valid Borel windows.
The work in Ref.~\cite{fesr} has closely viewed this problem 
and indeed they gave up relying on the Borel technique.

Our main objective of this paper is to put forward a solution 
to the problem in the BSR, by illustrating the $I=0$ and $J=1/2$ case of 
the $\Theta^+$, in a general way for exotic hadrons beyond the BSR. 
We here summarize the essential points of our analysis:
(I) In order to incorporate 
low-energy contributions more into our analysis,
we take into account 
the higher-dimensional terms of the OPE up to dimension 15. 
(II) 
Through
a favorable linear combination of correlation
functions, we suppress the 
high-energy
continuum contamination 
with the aid of the chiral symmetry in analogy of the Weinberg sum rule.
These technical developments enable us to 
establish Borel window wide enough to investigate the low energy hadronic
properties of the resonance and scattering states.  
 
This paper is organized as follows.
In Sec.~II, we briefly review the basic concepts of QSR's 
with special emphasis on the importance of
the pole dominance and the higher order terms in the OPE,
discussing the problems in the previous works.
In Sec.~III, to overcome the problems in previous works, 
we introduce a linear combination of 
the correlation functions with the aim of suppressing the continuum 
contamination, in which chiral symmetry plays an important role
in this cancellation.
We also discuss calculation of the OPE
and show all the OPE terms used in our analysis.
In Sec.~IV, we show our Borel analysis focusing on the criterion 
to set up the Borel window. We confirm the pole dominance
and the OPE convergence. The values of the $\Theta^{+}$ mass and 
residue obtained in this analysis are also shown. 
Sec.~V is devoted to a brief discussion on the $KN$ scattering states
and on relation between experimental observation and
our correlation function analysis.
Finally we summarize this work in Sec.~VI.

\section{The basic concepts of QSR
and problems in the previous works}
Following the standard way of the QCD sum rule, 
we start with the time-ordered two-point correlation function
defined by
\begin{eqnarray}
i\int d^{4}x e^{iq\cdot x} \langle 0 | T[J(x) \bar J(0)] | 0 \rangle
=  \hat q\Pi_{0} (q^{2}) + \Pi_{1} (q^{2}),
\end{eqnarray}
where $\hat q \equiv q^{\mu} \gamma_{\mu}$ and $\Pi_{0,1}(q^{2})$ 
are called the chiral-even and odd parts, respectively. 
Here $\langle 0|\cdots|0 \rangle$ denotes a vacuum expectation value 
(hereafter for brevity $\langle \cdots\rangle$).
The interpolating field $J(x)$ for the $\Theta^{+}$ 
consists of five quark fields with its quantum number.
The QSR is then obtained 
through the dispersion relation,
\begin{equation}
{\rm Re}\Pi_i(q^2) = P\int_0^{\infty}ds\,
[{\rm Im}\Pi_i(s)/\pi]/(s-q^2)  
\label{eq:disp}
\end{equation}
for $i=0,1$. 
${\rm Im}\Pi_i(s)$ 
satisfies the spectral conditions,
\begin{eqnarray}
\label{condition}
{\rm Im}\Pi_0(s) \geq 0,\  
\sqrt{s}{\rm Im}\Pi_0(s) - {\rm Im}\Pi_1(s) \geq 0.
\end{eqnarray} 
For sufficiently large $-q^2$, 
the left hand side of (\ref{eq:disp}) can be expressed by the OPE
with products $C_{i}$ of the Wilson coefficients and the vacuum condensates:
\begin{eqnarray}
\Pi_i^{ope}(q^2)
= \sum_{j=0}^{5}C_{2j+i}\ (q^2)^{5-j}\log(-q^2)
+ \sum_{j=1}^\infty\frac{C_{10+2j+i}}{(q^2)^j}.\,
\label{ope_}
\end{eqnarray}
The OPE starts from $(q^{2})^{5} \log (-q^{2})$ 
reflecting the large number of quark fields
in the $\Theta^{+}$ interpolating field.

The imaginary part in the right hand side of (\ref{eq:disp}) 
is parameterized as the hadronic spectrum.
We use the conventional pole plus continuum spectrum
with a single $\Theta^+$ resonance:
\begin{eqnarray}
{\rm Im} \Pi_i^{h}(s) \simeq \pi \lambda_i^2\delta(s-m_{\Theta^+}^2) 
+ \theta(s-s_{th}) {\rm Im} \Pi_i^{ope}(s).
\end{eqnarray}
Here $m_{\Theta^+}$ denotes the $\Theta^+$ mass, and
the residue $\lambda_i$ is the coupling strength of 
interpolating field to the resonance, satisfying 
\begin{eqnarray}
\pm m_{\Theta^{+}}\lambda_{0}^2
=\lambda_{1}^2
\end{eqnarray}
for parity $\pm 1$ respectively.
The second term represents
a model of continuum contribution with its threshold $s_{th}$
based on the simple duality ansatz. 
The QSR (\ref{eq:disp}) gives the physical quantities 
($m_{\Theta^+}$, $\lambda_i$) in terms of the known QCD parameters 
appearing in Eq.~(\ref{ope_}) \cite{shifman}.

Our QSR's are obtained 
from Eqs.~(\ref{eq:disp}) and (\ref{ope_})
for the chiral even and odd parts independently,
by using the Borel transformation technique \cite{reinders}, 
which qualitatively improves isolation of the $\Theta^{+}$ pole:
\begin{eqnarray}
\lefteqn{\lambda_i^2 e^{-m_{\Theta^+}^2/M^2}
= \sum_{j=1}^\infty\frac{(-)^j}{\Gamma(j)}\
\frac{C_{10+2j+i}}{(M^2)^{j-1}} } \ \ \ \ \  \ \
&& \nonumber \\
&&+\left(\int_0^\infty - \int_{s_{th}}^\infty\right) ds
 \, e^{-s/M^2} \sum_{j=0}^{5}C_{2j+i}\ s^{5-j}
\label{sum rule2}
\end{eqnarray}
where the continuum term in the hadronic spectrum is transferred to
the second integral in the right hand side using the duality ansatz.
It is worth noting here that, in this ansatz, the continuum term
is expressed in terms of the logarithmic terms in Eq.~(\ref{ope_}) and it
appears in the Borel sum rules (\ref{sum rule2}) 
as the integral of $\exp(-s/M^{2})$ 
weighted by polynomials of $s$. 
For later convenience, we define
\begin{eqnarray}
A_i(M^2;s_{th}) &=& \int^{s_{th}}_{0} \!\! ds\ e^{-s/M^2}
\frac{1}{\pi}{\rm Im} \Pi_i^{ope}(s) \nonumber \\
 &=& \sum_{j=1}^\infty\frac{(-)^j}{\Gamma(j)}\
\frac{C_{10+2j+i}}{(M^2)^{j-1}} \nonumber \\
&&+\int_0^{s_{th}}ds
 \, e^{-s/M^2} \sum_{j=0}^{5}C_{2j+i}\ s^{5-j} \ ,
\label{pole1} \\
B_i(M^2;s_{th}) &\equiv& 
\int_{s_{th}}^\infty \!\! ds\ e^{-s/M^2}
\frac{1}{\pi}{\rm Im} \Pi_i^{ope}(s) \nonumber \\
&=& \int_{s_{th}}^\infty \!\! ds\ 
e^{-s/M^2} \sum_{j=0}^{5}C_{2j+i}\ s^{5-j}.
\label{pole2}
\end{eqnarray}
$A_i(M^2;s_{th})$ is equal to the right hand side of Eq.~(\ref{sum rule2}).
These functions give portions of the Borel integration 
of $\Pi_{i}^{ope}$ for given threshold $s_{th}$.

In the pole-plus-continuum ansatz,
the threshold $s_{th}$ is very important parameter. It divides 
the hadronic spectral function into two parts:
\begin{eqnarray}
{\rm Im} \Pi_i^{h}(s) = \theta(s_{th}-s)\,{\rm Im}\Pi_i(s)
+ \theta(s-s_{th})\,{\rm Im}\Pi_i(s). \label{eq:divide}
\end{eqnarray}
The second term is approximated to the spectral function calculated
in the OPE, while the first term acts for the low-energy hadronic
contributions. In this sense, $s_{th}$ represents an energy scale 
where the quark-hadron duality ansatz works in the QSR analysis.
Thus, $s_{th}$ does not necessarily match a physical threshold
of the hadronic scattering states. Here
we assume the first term in Eq.~(\ref{eq:divide}) as a pole term.
This low-energy contribution, however, can contain both the hadronic 
resonance and scattering states below $s_{th}$. 
The validity of the pole assumption can be 
checked in the Borel stability analysis as discussed below.

The mass $m_{\Theta^+}$ is obtained by logarithmic derivative of 
Eq.~(\ref{sum rule2}) as 
\begin{eqnarray}
m_{\Theta^+}^2(M^2;s_{th})=d\log A_i/d(-1/M^2).
\end{eqnarray}
The pole residue is calculated together with 
the mass obtained above as 
\begin{eqnarray}
\lambda_i^2(M^2;s_{th})= A_i\exp\left[m_{\Theta^+}^2(M^2;s_{th})/M^2\right].
\end{eqnarray}
The two physical quantities should be, in principle, independent of 
the artificially introduced Borel mass $M$ within the ``valid'' Borel window 
that allows us reliable
extraction of the hadron property from this analysis.
The lower and higher boundaries of $M^2$
are determined from the 
OPE convergence and
the pole dominance, respectively
(see Sec.~IV in detail).
The pole dominance means that the low-energy first term 
in Eq.~(\ref{eq:divide}) is superior to the high-energy second 
term. This is necessary to extract the 
low-energy contribution from the integral
of the correlation function.
The threshold parameter $s_{th}$ 
should be also determined in the Borel analysis 
so as to make the mass and 
residue most insensitive to the change of $M^2$.
Therefore, all the physical quantities,
such as the mass, residue 
and continuum threshold,
are determined within the Borel analysis without any control parameters.

Now let us explain the problem in the pentaquark QSR. 
In the QSR analysis, the pole dominance to the continuum contribution
in the spectral function is essential to extract the desirable pole 
information. In the $\Theta^+$ case, however, 
the continuum contributions are potentially large, since the logarithmic terms
appear widely up to higher orders of the OPE. This stems from 
the higher mass dimension of the correlation function than the ordinary 
baryon case owing to the larger number of quark fields in the $\Theta^{+}$ 
interpolating field.
Consequently, despite improvement of the Borel transformation~\cite{reinders} 
to Eq.~(\ref{eq:disp}),
it is hard to establish the pole dominance in the spectral function, and
thus this makes the prediction of the $\Theta^+$ mass much less reliable.
In fact the prior pentaquark BSR's~\cite{SDO,bsr,Oga} evaluated the 
mass under a
condition of small pole contribution ($\lesssim 20\%$), 
as claimed in Ref.~\cite{fesr}.
In the BSR the magnitude of the continuum suppression can be measured 
by checking the Borel window, within which one searches for the Borel 
stability.
When obtaining better suppression, one may have a wider
window if the OPE convergence is also realized.

It is worth mentioning
the importance of the higher-dimensional terms in the OPE and the pole 
dominance.
When one neglects the higher-dimensional terms and/or the pole dominance,
one cannot establish the Borel window enabling reliable extraction
of the physical quantities, 
and also would encounter ``artificial'' Borel stability,
which is independent of the threshold parameter. 
Then the threshold is merely an adjustable parameter 
to reproduce the other physical quantities such as the mass and residue.
But if one includes the higher-dimensional terms
and establishes the pole dominance, then the threshold parameter is not 
adjustable any more but has a meaningful role for
stabilizing the physical quantities
within the Borel window for the change of Borel mass.
This is also the case 
even in the $\rho$-meson sum rule, where 
we indeed need inclusion of the dimension 6 terms in the OPE to avoid
the ``artificial'' stability, so that we can obtain 
$\rho$-meson mass close to the experimental data.

\section{Calculation of the linear combination of the correlators}
\subsection{Linear combination of the correlators}
To overcome the problems discussed in the previous section,
 {\it i.e.}\ to find out
true pole-dominance from the ${\Theta^+}$ correlation function, 
we propose a new setup of the ${\Theta^+}$ correlation function which
couples to less continuum states with the help of chirality 
of the interpolating fields.
This idea is to make use of an interesting property 
in the Weinberg spectral function sum rule \cite{Weinberg},
where the unlike chirality combination of the vector and axial-vector 
correlators, $\langle V_{\mu}(x) V_{\nu}(0) - A_{\mu}(x) A_{\nu}(0)\rangle$, 
vanishes in the limit of $x\rightarrow 0$.
This means that leading-orders are suppressed in the OPE.  
As favorable, it inevitably requires that one takes into account 
higher-dimensional operators which reflect the low-energy
physics, beyond the logarithmic terms. 
Let us consider the following interpolating fields with $I=0$ and $J=1/2$
based on the diquark picture \cite{JW}:
\begin{eqnarray}
P = \epsilon^{cfg} Q^c Q^f_{\mu 5} \gamma^{\mu} C \overline{s}^T_g,\ \ \
S = \epsilon^{cfg} Q_5^c Q^f_{\mu 5} \gamma^{\mu} \gamma_5 C 
\overline{s}^T_g,
\label{eqn:current}
\end{eqnarray}
where the diquark operators are defined by
\begin{eqnarray} 
Q^c&=&\epsilon^{abc} \{u^T_a C d_b\}, \\
Q^c_5&=&\epsilon^{abc} \{u^T_a C \gamma_5 d_b\},\\
Q^c_{\mu 5}&=&\epsilon^{abc} \{u^T_a C \gamma_{\mu} \gamma_5 d_b\}.
\end{eqnarray}
having the Lorentz covariant pseudoscalar, scalar and vector structures, 
respectively, with color indices $a,b,c,\cdots$,
the charge conjugation matrix $C$ 
and the transpose $T$~\cite{Sasaki:2003gi}.
Note that these interpolating fields definitely have $J=1/2$ due
to the $\gamma_{\mu}$ acting on $\bar{s}^{T}$~\cite{chung}.

In the above construction of the interpolating field, 
the pseudoscalar and scalar diquarks, 
$Q^a$ and $Q^a_5$, have been introduced into $P$ and $S$, respectively.
The essential point to reduce the leading orders of the OPE is that
the linear combinations between these diquarks, 
$Q^{a} \pm Q_{5}^{a}$, have the opposite chirality each other \cite{Jido}.
When one takes a relevant linear combination between the correlators 
with such an opposite chirality, the leading-orders suppression takes
place 
in the same way as the Weinberg sum rule for the vector currents. 
Motivated by this observation, 
we consider the following linear combinations of two correlators 
$P(x)\bar{P}(0)$ and $S(x)\bar{S}(0)$ with a mixing parameter $t$:
\begin{eqnarray}
\lefteqn{i\int d^4x \, e^{iq\cdot x}
\left\langle 0 \left|T [P(x)\bar{P}(0) 
- t\, S(x)\bar{S}(0)] \right|0 \right\rangle}
\nonumber\\
&\equiv& [\Pi_0^P(q^2) - t\, \Pi_0^S(q^2)]\, \hat{q}
+[\Pi_1^P(q^2) + t\,\Pi_1^S(q^2)], 
\label{eqn:correlation}
\end{eqnarray}
where
$\Pi_{0(1)}^{P[S]}$ are chiral-even (odd) parts of 
the correlation function for the currents $P$ $[S]$. 
The leading-orders suppression in the OPE is realized in $t=1$
and $t=-1$. In the former (later) case, the OPE starts from 
dimension 6 (7) in the chiral even (odd) part.
(See Eq.~(\ref{ope}) for the explicit OPE forms). 
The mixing parameter $t$ will be fixed according to the Ioffe's
optimization criteria \cite{IoffeZ}, that is, the sum rules satisfy 
sufficient continuum suppression and OPE convergence
at the same time. 
\subsection{The results of the OPE calculation}
Our strategy for the OPE calculation is as follows:
(I) We calculate the OPE up to dimension 15, which is higher enough
than the maximum dimension in logarithmic terms.
It is worth mentioning that the pentaquark currents may give extremely
slow OPE convergence in higher dimensions than six, since
the creation of a quark condensate by cutting loops 
costs a large factor, such as  $(4\pi)^2$. 
On the other hand, higher terms than dimension 12 are qualitatively
less important, because one can no longer diminish loops
by cutting hard quark lines due to the momentum 
conservation \cite{shifman}.
(II) We disregard radiative loop corrections.
These will be important in low dimensions
like the logarithmic terms [19], but in our analysis such
logarithmic terms are largely suppressed.
(III) The dependence of strange quark mass $m_s$ is evaluated to $O(m_s)$.
(IV) The higher-dimensional gluon condensates such as the triple gluon 
condensates are also neglected, because they are expected
to be smaller than the quark condensates entering 
in the tree-diagrams \cite{IoffeN}.
(V) We make good use of the vacuum saturation 
\cite{shifman} and factorization hypotheses \cite{leinweber}
in order to estimate less-known values of the high-dimensional condensates
as products of the lower-dimensional condensates.
To account for an uncertainty arising from this approximation, 
we will later exhibit final results
with the moderate errors.

Based on the above strategy, we obtain the explicit form of 
$C_n$ in (\ref{sum rule2}),
summing up all terms of the same dimension $n$ 
and taking the linear combination with $t$:
\begin{widetext}
\vspace{-0.4cm}
\begin{eqnarray}
&C_0= \frac{1-t}{2^{16}\ 3^2\ 5^2\ 7 \pi^8} \ ,\ 
C_1=  -\frac{(1+t) m_s}{2^{17}\ 3^2\ 5^2 \pi^8} \ ,\ 
C_3= - \frac{(1+t) R_s a}{2^{16}\ 3^2\ 5 \pi^8}\ ,\  
C_4=  -\frac{ (1-t) m_s R_s a}{2^{14}\ 3^2\ 5 \pi^8}
   +\frac{ (1-t) b}{2^{20}\ 3^2 \pi^8} \
 ,\ \nonumber\\
&C_5= \frac{(1+t) m_1^2 R_s a}{2^{16}\ 3^2\pi^8} 
  + \frac{(1+t) m_s b}{2^{20}\ 3\pi^8}\
 ,\ 
C_6= \frac{ (1-t) m_s m_1^2 R_s a}{2^{15}\ 3^2\pi^8} 
  -  \frac{(3-t) a^2}{2^{14}\ 3^2 \pi^8}\ , \
C_7= \frac{(1+t) R_s ab}{2^{18}\ 3^3 \pi^8} 
   +\frac{(3-t) m_s a^2 }{2^{12}\ 3^2\pi^8} \ , \nonumber \\ 
&C_8= - \frac{ 5(1-t) m_s R_s ab}
    {2^{18}\ 3^2 \pi^8} 
   +\frac{(75-23t) m_0^2 a^2 }
 {2^{16}\ 3^2\pi^8}\ , \
C_9= - \frac{(1+t) m_1^2 R_s ab}{2^{19}\pi^8}
+\frac{(3-t)\ R_s a^3}{2^{11}\ 3^2\pi^8}
-\frac{(9-4t)\ m_s m_0^2 a^2}{2^{14}\ 3\pi^8}\ , \nonumber \\
&C_{10}= \frac{ 5(1-t) m_s m_1^2 R_s ab}{2^{19}\ 3^2\pi^8}
+ \frac{ (3-t) m_s R_s a^3}{2^{12}\ 3^2\pi^8} 
- \frac{(337 - 121 t) m_0^4 a^2}{2^{19}\ 3^2\pi^8}
- \frac{(141-23t) a^2 b}{2^{18}\ 3^3 \pi^8}\ , \nonumber \\
&C_{11}= -\frac{(3-t) m_1^2 R_s a^3}{2^{12}\ 3^2\pi^8}
  - \frac{(9-4t) R_s m_0^2 a^3 }{2^{13}\ 3^2 \pi^8 } 
  + \frac{(19-18t) m_s m_0^4 a^2}{2^{16}\ 3^2 \pi^8 }
  + \frac{(99+67t)\ m_s a^2 b}{2^{17}\ 3^3 \pi^8}\ , \nonumber \\
&C_{12}= - \frac{(3-t) m_s m_1^2 R_s a^3 }{2^{12}\ 3^3\pi^8}
  -\frac{(17-4t)m_s m_0^2 R_s a^3}{2^{14}\ 3^2 \pi^8 } 
 +\frac{(23-33t) m_0^2 a^2 b}{2^{18}\ 3^3 \pi^8 } 
+\frac{(1+t) a^4}{2^{10}\ 3^3 \pi^8}\ , \nonumber \\
&C_{13}= \frac{(49+17t) R_s a^3 b}{2^{17}\ 3^3 \pi^8}
  + \frac{(9-4t) m_0^2 m_1^2 R_s a^3 }{2^{15}\ 3^2\pi^8 }
  + \frac{(19-18t) m_0^4 R_s a^3 }{2^{16}\ 3^3\pi^8 }
  + \frac{(1-11t) m_s m_0^2 a^2 b}{2^{18}\ 3\pi^8 }
  - \frac{(1-t) m_s a^4}{2^9\ 3^3 \pi^8 }\ , \nonumber \\
&C_{14}= - \frac{(17-4t) m_s m_0^2 m_1^2 R_s a^3 }{2^{15}\ 3^3\pi^8}
  - \frac{(67-18t) m_s m_0^4 R_s a^3 }{2^{17}\ 3^3 \pi^8 }
  - \frac{ (99-65t) m_s R_s a^3 b}{2^{18}\ 3^4 \pi^8} 
  + \frac{13(1+t) m_0^2 a^4}{2^{15}\ 3^2 \pi^8}\ ,\ 
C_{15}= \frac{13 (1-t) \frac{\alpha_s}{\pi} R_s a^5}{2^{10}\ 3^4 \pi^8},
\label{ope}
\end{eqnarray}
\end{widetext}
where $a=-(2\pi)^2\langle \bar{q}q \rangle$,
$b=(2\pi)^2\langle (\alpha_s/\pi)\, G^2 \rangle$,
$R_s=\langle \bar{s}s \rangle/\langle \bar{q}q \rangle$,
$m_0^2=\langle \bar{q}g_s\sigma\cdot Gq \rangle/\langle \bar{q}q \rangle$,
and $m_1^2=\langle \bar{s}g_s\sigma\cdot G s \rangle/\langle \bar{s}s \rangle$ 
with $q=u,d$ and strong coupling $\alpha_s=g_s^2/(4\pi)$.

The values of the QCD parameters are taken as 
$\langle (\alpha_s/\pi)\, G^2 \rangle = 0.0127\pm0.02$ GeV$^4$,
$m_s=0.12$ GeV, $\alpha_s(1\,{\rm GeV})=0.3$, $m_0^2=m_1^2=0.8\pm0.1 $GeV$^2$,
$R_s=0.8$ and 
$\langle \bar{q}q \rangle =-(0.230\pm0.020\, 
{\rm GeV})^3$~\cite{shifman,reinders}.
At first we use the central values,
and then we discuss the dependence of these uncertainties on our results. 
The recent work~\cite{Oga}, which only focuses on our correlator with $t=0$,
does not consider 
all terms of 
the quark-gluon mixed condensate $\langle \bar{q}g_s\sigma\cdot G q\rangle$
consisting of 
$\bar q q$ on a quark line and 
a soft gluon $G$ emitted from another quark line. We find that these terms are 
so important as to give $20\sim 30$\% contribution in dimension 8.

\section{Borel analysis}
\subsection{The criterion for the Borel window}
The Borel window in our analysis is determined as follows based on
Ref.~\cite{reinders}:
The lower boundary of the window is set up so as to make
the OPE convergence sufficient in higher-dimensional operators.
The criterion is 
quantified so that the highest-dimensional terms in the truncated OPE are 
less than 10\% of its whole OPE. At the same time, the higher boundary
of the window is fixed by the pole-dominance condition
that 
\begin{eqnarray}
\label{eq:conpole}
C_i(M^2;s_{th}) \equiv
\frac{|A_i(M^2;s_{th})|}{|A_i(M^2;s_{th})|
+|B_i(M^2;s_{th})|}
\ \gsim 0.5,
\end{eqnarray}
where $A_i$ and $B_i$
represents the pole and continuum contributions
defined in Eqs.~(\ref{pole1}), (\ref{pole2}) respectively.
The reason why we take the absolute values is 
that the continuum contribution can be no longer positive
definite in some regions of $M^2$, due to taking the linear combination 
of the correlation functions. 
This is a stronger condition than the criterion in Ref.~\cite{reinders},
where they do not take the absolute values of the correlation function.
Note that the 50\% pole contribution in our criterion is extremely 
large in comparison with any prior pentaquark sum rules, 
where the pole contributions are not more than 20\%
in their moderate windows.
Our conditions also satisfy the Ioffe's criteria to good accuracy.

\subsection{The pole dominance}
\begin{figure}[tb]
    \begin{minipage}{0.25\linewidth}
     \begin{center}
      \hspace{-2.0cm}
     \includegraphics[width=4.0cm, height=3.3cm]{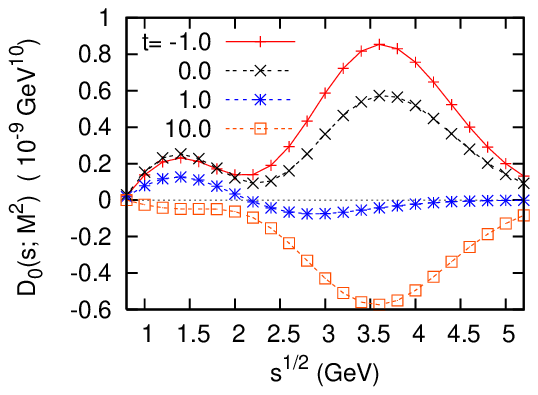}
     \end{center}
  \end{minipage}
\hspace{0.0cm}
   \begin{minipage}{0.25\linewidth}
    \begin{center}
     \hspace{-2.0cm}
     \includegraphics[width=4.1cm, height=3.3cm]{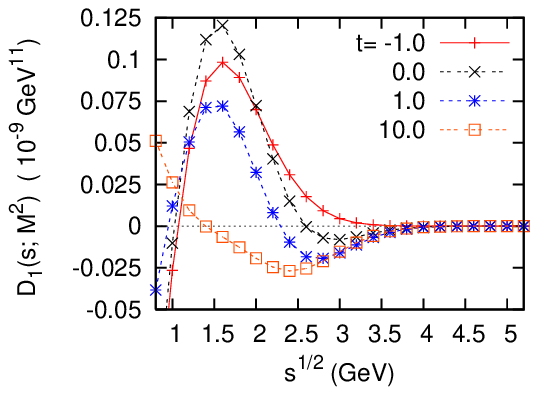}
      \end{center}
  \end{minipage}
\vspace{-0.1cm}
\caption{\label{fig:cont1} (Color online)
 The $t$-dependence of the behavior of 
$D_i(s;M^2)$ 
as a function of $\sqrt{s}$. The left panel is shown for the even part ($i=0$)
at $M^2=2.5$ GeV$^2$ and the right for the odd part
($i=1$) at $M^2=1.2$ GeV$^2$.}
\vspace{-0.4cm}
\end{figure}

First we check the continuum suppression of the chiral even (odd) 
correlation function in the case of $t=1$ $(t=-1)$.
In order to see the suppression of the high-energy contribution 
qualitatively, in Fig.~\ref{fig:cont1} we show
the behavior of the integrands in the right hand side 
of Eq.~(\ref{sum rule2}) normalized by $ \sqrt{1+t^2} $, {\it i.e.}
\begin{eqnarray}
D_i(s;M^2) \equiv \frac{ e^{-s/M^2} }{\pi \sqrt{1+t^2} }
\ {\rm Im} \Pi_i^{ope}(s),
\end{eqnarray}
which appears to the QSR as the integrands of continuum contributions.
These functions are plotted as a function of $\sqrt{s}$ with 
$t= -1.0,\ 0.0,\ 1.0,\ 10.0$,
which correspond to the
$P\overline{P}+S\overline{S}$, $P\overline{P}$,
$P\overline{P}-S\overline{S}$ and $S\overline{S}$-dominant cases 
in (\ref{eqn:correlation}), respectively.
Here we fix the Borel mass $M^{2}$ to be $2.5$ GeV$^2$ (even)
and $1.2$ GeV$^2$ (odd), which are in the Borel windows as discussed
later.
Note that in Fig.1 we plot
the combinations of the spectral functions, 
not the spectral functions themselves. 
As we remarked in Eq.~(\ref{condition}),
each correlation function does satisfy the spectral conditions. 
Then the linear combinations of correlation functions do not 
need to satisfy the conditions any more.

The left panel for the even part shows that the integrand with $t=1$ is 
successfully suppressed at $\sqrt{s}\gtrsim 2$ GeV, 
while in the other cases there are large contributions 
at $\sqrt{s}\sim 3$ GeV. 
On the other hand, in the odd part,
the best continuum suppression 
takes place at $\sqrt{s}\gtrsim 3$ GeV for $t=-1$. However, 
the contributions in the intermediate 
energies ($\sqrt{s}=2\sim 3$ GeV) are still so large 
that the isolation of the 
pole contribution is inadequate.
Similar tendency is also seen for $t=0,1$ except $t=10$, 
where we can no longer ensure
the pole dominance at moderate energy. 
Instead, we will take an optimal $t$ giving good OPE convergence.
This allows us to use lower Borel masses, where the Borel weight
leads to larger continuum suppression.

\begin{figure}[ht]
\begin{center}
          \includegraphics[width=8.6cm]{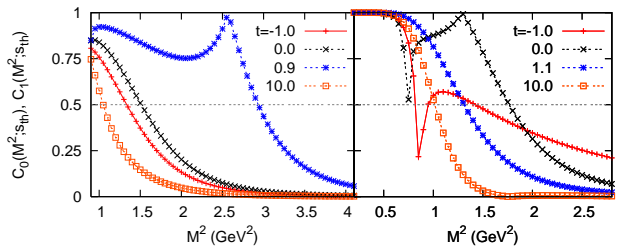}
     \end{center}
\caption{\label{fig:ope3} (Color online) The behavior of 
$ C_i(M^2;s_{th}) $
as a function of $M^2$.
 The left panel is shown for the even part
with $t=-1.0, 0.0, 0.9, 10.0$ at $\sqrt{s_{th}}=2.2$ GeV and 
the right for the odd part
with $t=-1.0, 0.0, 1.1, 10.0$ at $ \sqrt{s_{th}}=2.1$ GeV.}
\end{figure}

The pole dominance can be checked also in Fig.~\ref{fig:ope3}, where
we plot $C_i(M^2;s_{th})$ as functions 
of the Borel mass with a fixed $s_{th}$. 
The function $C_{i}$ defined in Eq.~(\ref{eq:conpole}) measures
the pole dominance and is used for the criterion of the Borel window
that the upper boundary is determined so that $C_{i} \ge 0.5$.
In the left panel we plot $C_0$ (even part) in the case 
of $t=-1.0,\ 0.0,\ 0.9,\ 10.0$
with $s_{th}$ fixed to $2.2$\ GeV,
and in the right panel we plot $C_1$ (odd part) in the case 
of $t=-1.0,\ 0.0,\ 1.1,\ 10.0$
with $s_{th}$ fixed to $2.1$\ GeV.
We find that, in the even part $C_{0}$ with $t=0.9$ and
in  the odd part $C_{0}$ with $t=0.0$, $1.1$, the pole contribution
dominates the correlation function over the wide range of the Borel mass. 
The appearance of cusp structures in $C_0$
over the range of $M^2=2.5\sim 3.0$\ GeV$^2$ at $t=0.9$, 
arises as the result of large cancellation of the continuum contribution 
in the denominator of $C_i$.

\subsection{The OPE convergence}
\begin{figure}[t]
\vspace{-0.0cm}
\begin{center}
          \includegraphics[width=10.8cm, height=7.0cm]{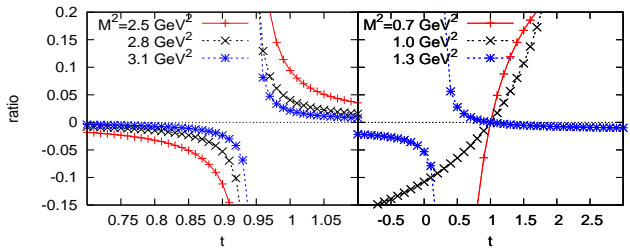}
     \end{center}
\vspace{-4.0cm}
\caption{\label{fig:ope} (Color online) The behavior of 
highest-dimensional terms of the OPE divided by the whole OPE
as a function of $t$
for various $M^2$. The left panel is shown for the even part
with $M^2=2.5,\ 2.8,\ 3.1$ GeV$^2$ and the right for the odd part
with $M^2=0.7,\ 1.0,\ 1.3$ GeV$^2$.}
\end{figure}

Next we discuss the OPE convergence, which determines the lower boundary 
of the Borel windows. 
Shown in Fig.~\ref{fig:ope} are the ratios of highest-dimensional terms 
(dimension 14 in the OPE for the even part and dimension 15 for the odd one)
to the whole OPE as a function of $t$ for various $M^2$.
Here we take the Borel mass $M^{2}$ to be $2.5, 2.8, 3.1$ GeV$^2$ (even)
and $0.7, 1.0, 1.3$ GeV$^2$ (odd), which are around the Borel windows 
set below. 
Our condition on the OPE convergence is that the ratios are less than 10\%.
For the even part having good continuum suppression at $t=1$, we
consider the vicinity of $t=1$.

In Fig.~\ref{fig:ope} we find quite good OPE convergence in the whole region 
except $t \simeq 0.95$. This remarkable convergence should 
be compared with the OPE of the current given in Ref.~\cite{SDO}, in which 
the convergence is not sufficient \cite{khj}.  
The exceptionally bad convergence at $t \simeq 0.95$ is due to 
cancellation in the whole OPE, which is rejected by our criterion for 
the Borel window. If it were not the case, 
we would need to take account of higher-dimensional terms truncated here. 
In the odd part, we investigate the OPE convergence 
in relatively low $M^2$-region
retaining the continuum suppression due to the Borel weight. 
The right panel of Fig.~\ref{fig:ope} shows that the OPE convergence 
is good around $t=1$, while bad around $t=-1$ where 
the continuum suppression is realized at higher $\sqrt{s}$.
Hence we take the mixing angle around $t=1$ 
in the odd part as well as in the even part.
\begin{figure}[ht]
\begin{center}
          \includegraphics[width=8.6cm]{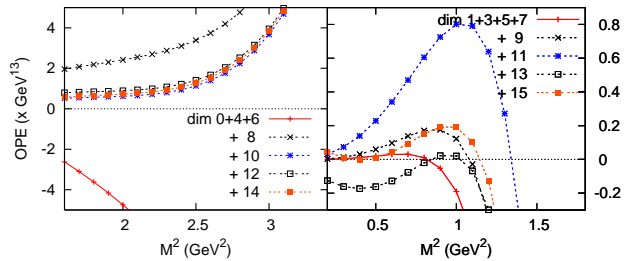}
     \end{center}
\vspace{-0.6cm}
\caption{\label{fig:ope1} (Color online)
OPE contributions added up each term in sequence.
The left panel is shown as a function of $M^2$
with $t=0.9$ fixed for the even part and the right with $t=1.1$ fixed
for the odd part.}
\end{figure}

To complement the issue of the OPE convergence, 
we illustrate OPE contributions as a function of $M^2$ with fixed $t$
in Fig.~\ref{fig:ope1}, where each dimension term is added up in sequence. 
We use $t=0.9$ 
for the even part (left panel) and $t=1.1$
for the odd part (right panel). 
It turns out that higher-dimensional terms become smaller for both parts.

\subsection{The Borel stability on the physical quantities}
After establishing the pole dominance and the OPE convergence,  
we move on setting the Borel window
and discussing the Borel stability on the physical quantities, i.e.
mass and residue.

Fine-tuning $t$ around $t=1$ to obtain the widest Borel windows,
we find our best Borel windows as
$2.5\le M^2 \le (2.9\sim 3.0)$ GeV$^2$ at $\sqrt{s_{th}}=2.1\sim 2.3$ GeV 
for the even part ($t=0.9$), and $0.7 \le M^2 \le (1.2\sim 1.3)$ GeV$^2$
at $\sqrt{s_{th}}=2.0\sim 2.2$ GeV for the odd part ($t=1.1$)
as seen from Fig.~\ref{fig:ope3}.
Here we have chosen the threshold parameters $\sqrt{s_{th}}$ so as to maximize
the correlation with the pole at $\sqrt{s}\lesssim 2$ GeV 
as roughly seen in Fig.~\ref{fig:cont1}. 
These thresholds indeed give better Borel
stability for the physical parameters.
\begin{figure}[b]
\vspace{-0.7cm}
\begin{center}
          \includegraphics[width=12.0cm, height=8.0cm]{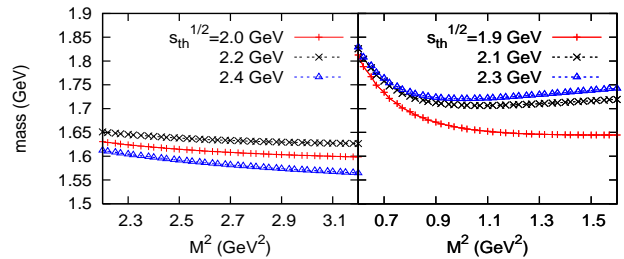}
     \end{center}
\vspace{-4.6cm}
\caption{\label{fig:mass} (Color online)
The $M^2$-dependence of the $\Theta^+$ mass for the even part (left) 
and for the odd part (right).}
\end{figure}
\begin{figure}[thb]
\vspace{-0.0cm}
\begin{center}
          \includegraphics[width=12.0cm, height=8.0cm]{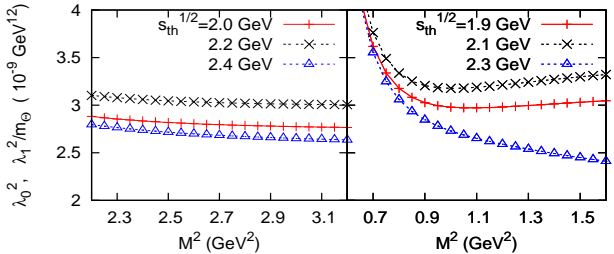}
     \end{center}
\vspace{-4.6cm}
\caption{\label{fig:resi} (Color online)
The $M^2$-dependence of the residues 
($\lambda_0^2$: left panel and $\lambda_1^2/m_\Theta$: right panel) 
in the same way with Fig.~\ref{fig:mass}. 
Note that the residue of the odd part is divided by $m_{\Theta^+}$.}
\vspace{-0.2cm}
\end{figure}
The values of the mass and the residue are evaluated
within the Borel windows determined above.  We plot
the  Borel mass dependence of the mass and the residue
in Figs.~\ref{fig:mass} and \ref{fig:resi}, respectively, where
the left (right) panel is the plot for the chiral even (odd) 
sum rule. Figure~\ref{fig:mass} shows that the Borel stability
is established quite well within the Borel windows. The best
stability is achieved with $\sqrt{s_{th}}= 2.2$ GeV (even) 
and $2.1$ GeV (odd), giving $m_{\Theta^+}= 1.64$ 
GeV (even) and $1.72$ GeV (odd) respectively. These masses
slightly depend on the change of the thresholds in the range of
$2.0\sim 2.4$ GeV (even) 
and 
$1.9\sim 2.3$ GeV (odd). 
With these uncertainties, the masses are evaluated as 
$m_{\Theta^+}=1.64\pm 0.03$ GeV (even) 
and 
$1.72\pm 0.05$ GeV (odd).
The residue is evaluated in the same way as the mass from the Borel
curve shown in Fig.~\ref{fig:resi}. We find quite good stability again. 
The values of the residue are obtained from the chiral even 
and odd sum rules as 
$\lambda_0^2 = (3.0 \pm 0.1)\times 10^{-9}$ GeV$^{12}$ 
and $\lambda_1^2/m_{\Theta^+} = (3.4 \pm 0.2) \times 10^{-9}$ GeV$^{12}$
respectively. 
It is remarkable that these numbers are quite similar with
the close $t$. 
This implies that our analysis investigates consistently the same state
in the two independent sum rules. 
Note that from the relative sign of the residue,
we assign {\it positive} parity to the observed $\Theta^+$ state.

We investigate the dependence of the QCD parameters on our final results.
We find that the final 
results are insensitive to the change of $m_s$, $R_s$, $m_1^2$, 
$\langle \bar{q}q \rangle$ and $\alpha_s$, and are marginally
sensitive to that of $m_0^2$ and $\langle (\alpha_s/\pi)\,G^2\rangle$.
By accounting for uncertainties of these QCD parameters 
and errors arising from our approximation,
we estimate the theoretical errors of our results to be totally around 15\%. 
Combining both 
results of the even and odd sum rules with this error, we finally 
conclude that our estimation of the $\Theta^{+}$ mass  
is $m_{\Theta^+}=1.68\pm 0.22$ GeV.

\begin{figure}[ht]
\begin{center}
          \includegraphics[width=8.6cm]{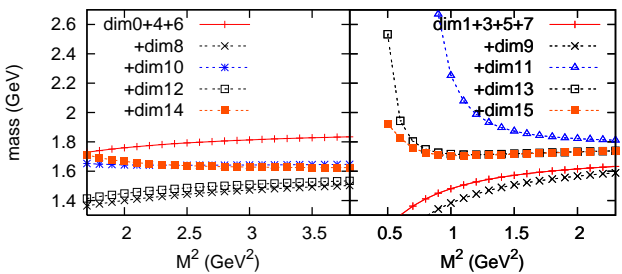}
     \end{center}
\vspace{-0.6cm}
\caption{\label{fig:ope2} (Color online)
The response of the $\Theta^+$ mass to OPE contributions
as a function of $M^2$, where each OPE term is added up 
as in Fig.~\ref{fig:ope1}. The left panel is shown for the even part
with $t=0.9, \sqrt{s_{th}}=2.2$ GeV fixed and 
the right for the odd part
with $t=1.1, \sqrt{s_{th}}=2.1$ GeV fixed.}
\end{figure}

We finally confirm the OPE convergence in the mass. 
In Fig.~\ref{fig:ope2},
we plot the response of the $\Theta^+$ mass to addition of the 
higher-order OPE contributions 
as a function of $M^2$ with $t=0.9, \sqrt{s_{th}}=2.2$ GeV for
the odd part (left panel) and $t=1.1, \sqrt{s_{th}}=2.1$ GeV 
for the odd part (right panel). 
We find that inclusion of higher OPE terms makes Borel curves 
more stable in both cases.

\section{Discussion}
Although we have not explicitly taken into account
the $KN$ scattering state appearing at about 100 MeV 
below the observed pentaquark mass,
in this section we make a brief comment on the contamination 
from this scattering state.
Efforts to isolate the considering $\Theta^+$ state 
from the scattering state in the analysis were recently 
made in \cite{su} for the QSR, 
and in \cite{lattice} for the lattice QCD.
Here we explain another aspect of such isolation in more 
intuitive and qualitative way.
The pole information of the $\Theta^+$ should be carried 
by higher-dimensional OPE terms beyond the suppressed lower-dimensional
terms which have more information about the perturbative region.
The property of the $\Theta^{+}$ in the QSR can be sensitive to
the values of the chiral condensate $\langle\bar{q}q\rangle$,
because the higher-dimensional
contribution is mainly controlled by the products 
of the chiral condensate $\langle\bar{q}q\rangle$. 
Our observed states for
both even and odd parts, however, are rather insensitive for
the change of the order parameter, while the $KN$ threshold 
is expected to be sensitive, since the nucleon mass is described
as the Ioffe's formula $M_N \propto (-\langle \bar{q}q\rangle)^{1/3}$
in the QSR \cite{reinders}.
Therefore, we speculate the possible isolation of the observed states 
from the $KN$ scattering state.
This would be also challenging for further investigating
the mass shift of the $\Theta^+$ for chiral restoration 
in matter \cite{navarra}
in comparison with the $KN$ scattering state.

Concerning the existence of the pentaquark state,
although the pronounced peak of the pentaquark has not been 
seen in experiments as typified by recent measurements at Jlab~\cite{clas}, 
it does not directly mean that our QSR calculations are incorrect.
We faithfully follow the original idea of QSR, 
especially emphasizing the importance of the pole dominance 
and the Borel stability, then we find a pentaquark state 
in our analysis. 
When comparing such a theoretical finding to the experimental 
observations, one needs further steps, for instance, 
consideration of the reaction mechanism.
The QSR just analyses spectral functions composed of resonance poles and
scattering states with interpolating fields prepared appropriately.
It may be the case that the ratio of the strengths between 
the pentaquark pole and the scattering states 
is different from those observed in the experiments. 
Our sum rule extracts successfully the pentaquark's pole contribution 
from the background scattering states, such as the $KN$ state, 
as well as the high-energy continuum contributions. 
This may indicate that the low-energy scattering states 
are also suppressed in our linear combination of 
the correlation functions.

\section{Summary}
In this work, we have presented a new idea to address exotic hadrons 
with a number of quarks, such as the pentaquark, in the QSR, 
where the favorable continuum suppression is realized 
by considering a linear combination of 
two correlators with different chirality.
Implementing the Borel technique, 
we indeed obtained the wide Borel windows 
that enable to extract hadronic properties much reliably from this analysis. 
We should bear in mind that as far as one relies on the simple
step-function 
form of continuum contribution and the duality ansatz, one could not 
easily construct any reliable BSR's without considering 
such continuum suppression.
With paying attention also to the OPE convergence, we finally estimate 
$m_{\Theta^+}=1.68\pm 0.22$ GeV including uncertainties of the condensates
in the OPE calculation up to dimension 15.
Here the Borel curves look fairly flat and almost independent 
of the continuum thresholds, 
and such a feature is also seen for the pole residue.

We would like to point out that 
choice of the interpolating fields solely
cannot achieve enough suppression of the large continuum contributions,
since the logarithmic terms of OPE give large continuum contributions
to the spectral function via the duality ansatz
due to the high-dimensional current of pentaquark.
To obtain sufficient continuum suppression,
it is important to take a favorable linear combination of the correlation
functions with the aid of chirality of the interpolating fields.
This idea would be also applicable for 
all the correlation function analyses as in lattice QCD,
where a contamination from the high-energy contributions
hinders extraction of information on low-energy hadron states,
and for other exotic hadrons like a tetraquark~\cite{Chen:2006hy}.
Also, it is noteworthy 
that a concept of chiral symmetry introduced here plays an important role
to pick up the information in the low-energy region.

\vspace{-0.5cm}
\begin{acknowledgments}
We thank Dr. T.T.~Takahashi for helpful discussion on their lattice
calculation. T.K. also acknowledges Prof. M. Asakawa for useful comments and
the members of Nuclear Theory Group at Osaka University for their hospitality.
\end{acknowledgments}

%
%

\end{document}